\documentclass[fleqn,10pt]{wlscirep}
\usepackage[utf8]{inputenc}
\usepackage[T1]{fontenc}
\usepackage{multirow}
\title{CT Image Segmentation for Inflamed and Fibrotic Lungs Using a Multi-Resolution Convolutional Neural Network}

\author[1,*]{Sarah E. Gerard, PhD}
\author[2]{Jacob Herrmann, PhD}
\author[3]{Yi Xin, MS}
\author[4]{Kevin T. Martin,BS}
\author[5,7]{Emanuele Rezoagli, MD, PhD}
\author[6]{Davide Ippolito, MD}
\author[5,7]{Giacomo Bellani, MD, PhD}
\author[4]{Maurizio Cereda, MD}
\author[1,8]{Junfeng Guo, PhD}
\author[1,8]{Eric A. Hoffman, PhD}
\author[9,1,8]{\\David W. Kaczka, MD, PhD}
\author[8,1]{Joseph M. Reinhardt, PhD}

\affil[1]{Department of Radiology, University of Iowa, Iowa City, IA, USA}
\affil[2]{Department of Biomedical Engineering, Boston University, Boston, MA, USA}
\affil[3]{Department of Radiology, University of Pennsylvania, Philadelphia, PA, USA}
\affil[4]{Department of Anesthesiology and Critical Care, University of Pennsylvania, Philadelphia, PA, USA}
\affil[5]{Department of Medicine and Surgery, University of Milano-Bicocca, Monza, Italy}
\affil[6]{Department of Diagnostic and Interventional Radiology, San Gerardo Hospital, Monza, Italy}
\affil[7]{Department of Emergency and Intensive Care, San Gerardo Hospital, Monza, Italy}
\affil[8]{Roy J. Carver Department of Biomedical Engineering, University of Iowa, Iowa City, IA, USA}
\affil[9]{Department of Anesthesia, University of Iowa, Iowa City, IA, USA}

\affil[*]{sarah-gerard@uiowa.edu}

\begin{abstract}

The purpose of this study was to develop a fully-automated
segmentation algorithm, robust to various density enhancing lung
abnormalities, to facilitate rapid quantitative analysis of computed tomography
images. 
A polymorphic training
approach is proposed, in which both specifically labeled left and
right lungs of humans with COPD, and nonspecifically labeled
lungs of animals with acute lung injury, were incorporated
into training a single neural network. The resulting network is intended
for predicting left and right lung regions in humans with or without
diffuse opacification and consolidation.  Performance of the
proposed lung segmentation algorithm was
extensively evaluated on CT scans of subjects with COPD, 
confirmed COVID-19, lung cancer, and IPF,
despite no labeled training data of the latter three diseases. 
Lobar segmentations were obtained using the left and right lung segmentation as input to the LobeNet algorithm. Regional lobar analysis was performed using hierarchical clustering to identify radiographic subtypes of COVID-19.
The proposed lung
segmentation algorithm was quantitatively evaluated using semi-automated and manually-corrected segmentations in 87 COVID-19 CT images, achieving an
average symmetric surface distance of $0.495\pm0.309$ mm and
Dice coefficient of $0.985\pm0.011$.  
Hierarchical
clustering identified four radiographical phenotypes of COVID-19
based on lobar fractions of consolidated and poorly aerated
tissue. Lower left and lower right lobes were consistently more
afflicted with poor aeration and consolidation. However, the most
severe cases demonstrated involvement of all lobes.
The polymorphic training approach was able to accurately segment
COVID-19 cases with diffuse consolidation without requiring
COVID-19 cases for training. 

\end{abstract}
\begin{document}

\flushbottom
\maketitle
%
%
\thispagestyle{empty}


\section*{Introduction}


Computed tomographic (CT) imaging has played an important role in
assessing parenchymal abnormalities in lung diseases such as chronic
obstructive pulmonary disease (COPD), and more recently, the novel
coronavirus disease (COVID-19). CT imaging is important for
diagnostics as well as quantifying disease involvement and progression
over time. CT-based disease quantification can
be used for patient stratification, management, and prognostication~\cite{zhou2020clinical,huang2020clinical}. 
Automated analysis of images is critical for objective
quantification and characterization of large numbers of CT datasets. 
In particular, reliable lung and lobe segmentation is an important precursor to
quantifying total lung and regional involvement of the disease.

Conventional lung and lobar segmentation approaches
programmatically achieve segmentation using prior information about voxel
intensity and second-order structure in small
neighborhoods~\cite{hu2001,van2010,kuhnigk2005,zhou2006,ukil2009a,
lassen2013,pu2009}. More advanced methods have used shape priors in the
form of atlases or statistical shape
models~\cite{sun2012,sofka2011,sluimer2005,zhang2006,van2009,pinzonlung}.
Recently, deep learning approaches have surpassed the performance
of rule-based segmentation by learning important
features for segmentation from labeled training data.  A multi-scale
CNN approach for segmentation of acutely injured lungs in animal
models demonstrated that incorporation of global features improved
lung segmentation in cases with diffuse
consolidation~\cite{gerard2020multi}.  FissureNet is a deep learning
based fissure segmentation method which identifies the boundaries
between lobes~\cite{gerard2018a}, a critical step for lobar
segmentation. Preliminary work on extending FissureNet to segment
lobes was proposed, although this method was only evaluated on
chronic obstructive pulmonary disease (COPD)
cases without density enhancing pathologies~\cite{gerard-isbi2019}. 
Other methods have directly learned lobe segmentation
without first explicitly identifying lungs and
fissures~\cite{george2017pathological,10.1007/978-3-030-00889-5_32}.

Automated lung segmentation in patients with COVID-19 is a challenging task, given
the multitude of nonspecific features that appear on CT (i.e.,
bilateral and peripheral ground-glass opacities and consolidation).
Intensity-based segmentation methods may fail to include infected
regions, which is critical for any image quantitative analysis. Furthermore,
lung opacities can obscure the fissure appearance, making it
challenging to identify lobes. CNNs have great potential for
automated segmentation due to their ability to identify low-level and abstract
features. However, a challenge with deployment of deep learning methods
in medical imaging is the accessibility to labeled training data
representative of all disease phenotypes - for example, a lobar segmentation
network trained only on data from COPD patients is unlikely to
perform well in COVID-19 patients with diffuse lung and focal lung consolidation.

\begin{figure}[ht]
\centerline{\includegraphics[width=\columnwidth]{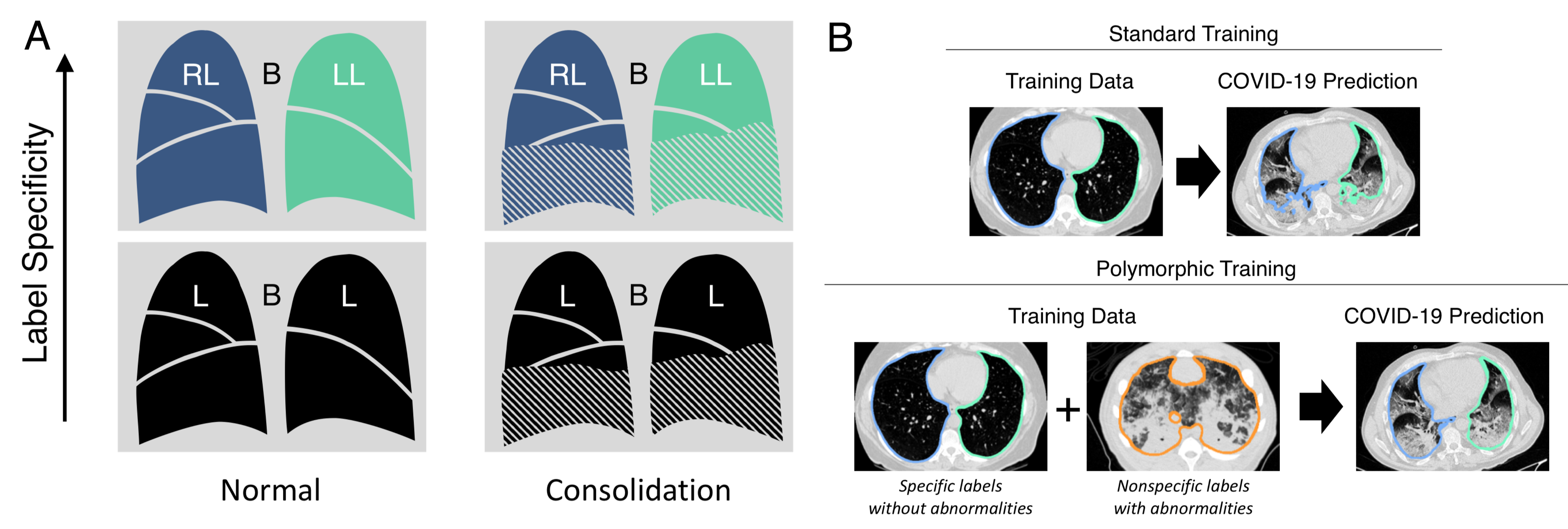}}
    \caption{Panel A: Motivation for polymorphic training. In this work, the
    desired segmentation target is consolidated cases with specific
    labels of left lung (LL), right lung (RL), and background (B)
    (upper right). However, only normal cases with specific labels
    (upper left) and consolidated cases with non-specific
    labels of lung (L) and background (B) (lower right) are
    available for training. The proposed polymorphic training approach
    allows us to utilize the available training data and generalize
    to the target domain of consolidated specifically labeled cases
    (upper right).
    Panel B: Standard training (top) using only specifically labeled COPD
    images lacks the consolidation phenotype necessary to successfully segment injured regions in COVID-19
    images. 
        Polymorphic training (bottom) utilizes specifically
    labeled COPD images with nonspecifically labeled animal models
    of acute lung injury to achieve specific lung labels including
    injured regions in COVID-19
    images. 
    The specific lung labels are depicted in green and blue
    for left and right lung, respectively. The nonspecific lung label
    is depicted in orange.}
\label{fig:poly}
\end{figure}

Additional labeled training data may be available,
although the labels may not have the desired level of specificity.
For example, a voxel corresponding to parenchymal tissue may simply be
labeled as lung (as opposed to non-lung) or it could be more
specifically labeled as left or right lung (see Figure~\ref{fig:poly}). Although nonspecific
labels may not be directly useful for training networks to predict
specific labels, the nonspecific dataset may still contain important
disease phenotypes absent from the dataset with specific labels.  We thus hypothesize
that data with generic labels can still be valuable when training a
network to predict specific labels.  Ideally, training would
accommodate labels with different degrees of specificity (i.e., a
hierarchical categorization). In this study, we propose a solution to accommodate
partially labeled training data, wherein ``partial'' refers to
different degrees of specificity in a hierarchical categorization of
labels. We refer to this solution as ``polymorphic'' training. Polymorphism in
biology and computer science refers to the ability of organisms and
data types to exist as one of multiple subtypes (e.g., schnauzer is a
subtype of dog, dog is a subtype of mammal). We propose a polymorphic
training strategy that injects supervision at different network
layers predicting different subtypes of voxel classification,
specifically for data with hierarchical labels.


The specific aim of this work was to develop an algorithm for
fully-automated and robust lung segmentation in CT scans of patients
with pulmonary manifestations of COVID-19, to facilitate regional
quantitative analysis.  In related work, FissureNet~\cite{gerard2018a}
and LobeNet~\cite{gerard-isbi2019} were proposed for robust segmentation of pulmonary fissures and lobes.
However, FissureNet and LobeNet cannot be applied directly to CT
images, but require an initial lung segmentation which distinguishes
left vs. right lung. Automated lung segmentation for COVID-19 images
is challenging due to diffuse consolidation obscuring lung
boundaries. In this work, we propose a segmentation method which
identifies left and right lungs in COVID-19 images. Given the
scarcity of labeled COVID-19 CT images available for training, two
existing datasets with complementary features were used: 1) a dataset
from patients with COPD, with specifically labeled left and right
lungs; and 2) a dataset from experimental animal models of acute lung
injury, with only a single nonspecific lung label.
The first dataset provides human training examples with specific left and
right lung labels, while the second dataset contains important
disease phenotypes (i.e., ground glass opacification and
consolidation) absent from the COPD images (see
Figure~\ref{fig:poly}). 
The design of the polymorphic training is motivated by a need to
accommodate labeled training data with heterogeneous degrees of
subclassification, since datasets may have a single label for all
lung tissue or labels distinguishing left and right lungs.

\section*{Materials and Methods}
\subsection*{Datasets}
The number of images used for training and evaluation are summarized in Table~\ref{tab:datasets}.
A combination of human and animal CT datasets with different diseases
were utilized for training the lung segmentation model. Human
datasets were acquired from COPDGene~\cite{regan2011}, a
multi-center clinical trial with over 10,000 COPD patients enrolled.
Animal datasets of acute lung injury models included canine, porcine, and ovine species
(see~\cite{gerard2020multi} for detailed description of datasets). In
total, 1000 human CT images and 452 animal CT images were
used for training the lung segmentation module. 
Note, only 1000 of the COPD CT images were used for training in effort to avoid a large imbalance between disease phenotypes in the training data.
All training CT images have a ground truth lung
segmentation generated
automatically using the Pulmonary Analysis Software Suite (PASS,
University of Iowa Advanced Pulmonary Physiomic Imaging
Laboratory~\cite{guo2008a}) with manual correction if
necessary.  For human datasets, ground truth segmentations
distinguished the left and right lungs, whereas the animal datasets had only
a single label for all lung tissue.  It is important to note that
separation of left and right lungs is not trivial due to close proximity
of the left and right lungs, especially in the three animal species used due to the
accessory lobe adjacent to both the left and right lungs. 

A dataset of 133 clinical CT images of COVID-19 patients was acquired
from: the Hospital of San Gerardo, Italy; University of Milan-Bicocca,
Italy; Kyungpook National University School of Medicine, South Korea; and Seoul National University Hospital, South Korea.
Patients were included based on confirmed COVID-19 diagnosis by nucleic acid amplification tests.
Data use was approved by Institutional Review Boards at University of Milano-Bicocca, the Hospital of San Gerardo, Kyungpook National University School of Medicine, and Seoul National University Hospital. Given the retrospective nature of the study and in the presence of technical difficult in obtaining an informed consent of patients in this period of pandemic emergency, informed consent was be waived and all data was anonymized.
All procedures were followed in accordance with the relevant guidelines.
Details from the Korean COVID-19 cases are provided in Nagpal et al~\cite{nagpal2020imaging}.
Ground truth lung
segmentations were performed for 87 cases using PASS~\cite{guo2008a} or pulmonary
toolkit (PTK)~\cite{doel2017pulmonary} with manual correction as
necessary.  Manual correction required an average of $94\pm48$
minutes per case.

To evaluate the performance on other pulmonary diseases, three
additional evaluation datasets were utilized: 5986 CT images
from COPDGene, 1620 CT images from lung cancer patients undergoing
radiation therapy, and 305 CT
images from patients with idiopathic pulmonary fibrosis (IPF). Ground
truth segmentations were generated using PASS followed by manual
correction.

 \begin{table}
     \centering
     \caption{Number of 3D CT images used for training and evaluation.}
     \begin{tabular}{lrr}
\toprule
{}          &  Training &  Evaluation \\
\midrule
COPDGene    &      1000 &     5986 \\
Animal ARDS &       453 &        - \\
Cancer      &         - &     1620 \\
IPF         &         - &      305 \\
COVID-19    &         - &       87 \\
\midrule
Total       &      1453 &     7998 \\
\bottomrule
\end{tabular}

     \label{tab:datasets}
 \end{table}

\subsection*{Multi-Resolution Model}
The LungNet module used a multi-resolution approach adapted
from~\cite{gerard2020multi} to facilitate learning both global and local
features important for lung segmentation. LungNet consists
of a cascade of two CNN models; the low-resolution model LungNet-LR
and the high-resolution model LungNet-HR.

LungNet-LR was trained using low-resolution images. All CT images and
target label images are downsampled to 4 mm isotropic voxels using
b-spline and nearest-neighbor interpolation for the CT and label images,
respectively. A Gaussian filter was applied to the CT images prior to
downsampling to avoid aliasing.  LungNet-LR yields a
three-channel image, corresponding to predicted probabilities for left
lung, right lung, and background. 

LungNet-HR was trained with high-resolution images. The
CT image, the output of LungNet-LR, and the target label image were
resampled to have 1 mm isotropic voxels for consistency. The CT image
and left/right probability maps were then combined to produce a three-channel input
for training the high-resolution network. Similar to LungNet-LR, the
output of LungNet-HR was a three-channel probability image. The final
lung segmentation was obtained by thresholding the left and
right probability channels at $p=0.5$.

\subsection*{Polymorphic Training}
\begin{figure}[ht]
\centerline{\includegraphics[width=\columnwidth]{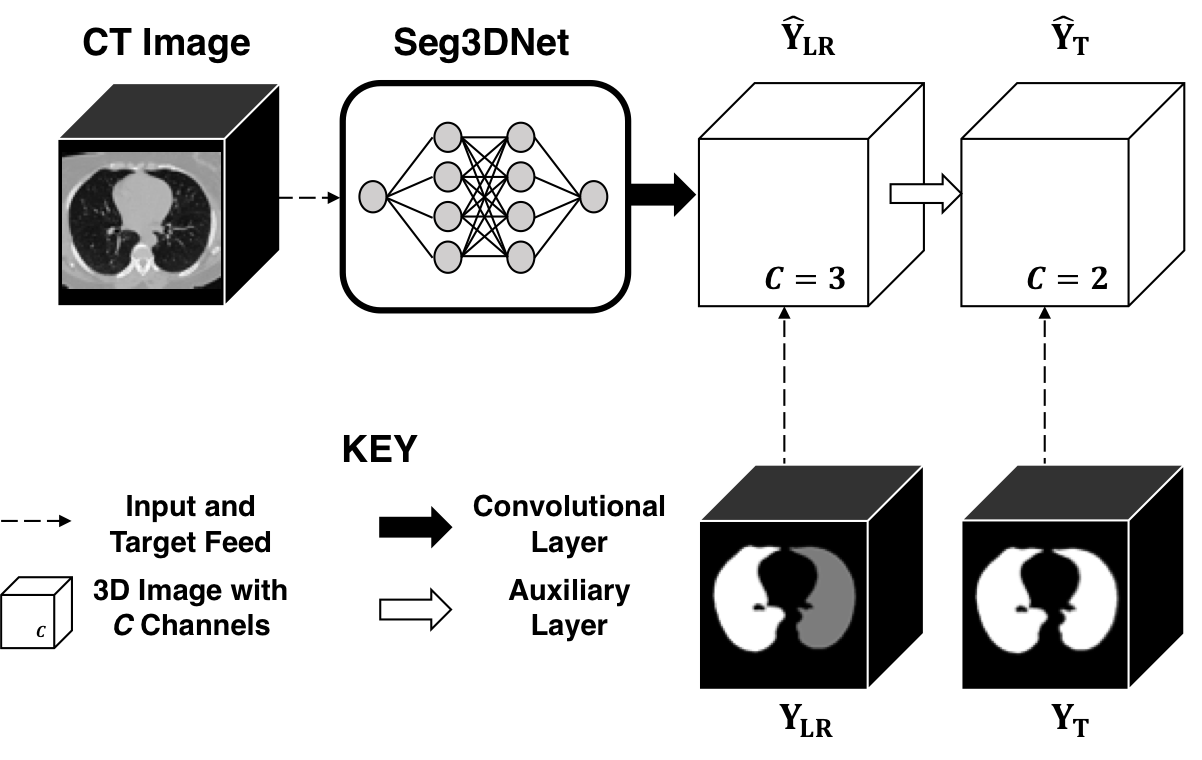}}
    \caption{Polymorphic training accommodates labeled data
    with different degrees of specificity. In this case some labeled
    training have specific labels distinguishing left and right lung,
    while other training data only have a single label for all lung
    tissue.}
\label{multiloss}
\end{figure}

We used a novel polymorphic training strategy, illustrated in
Figure~\ref{multiloss}, which incorporated all information in
partially labeled datasets. The ultimate goal was to train a network
that could distinguish left vs. right lung, with or without
abnormal pathological features. The three-channel prediction image
produced by the last layer of Seg3DNet,
denoted $\mathrm{\hat{Y}_{LR}}$, yielded channels corresponding to
left lung, right lung, and background probabilities. To make this
output compatible with the animal datasets, which have only a
single lung label, an auxiliary layer with
supervision was added to the network after $\mathrm{\hat{Y}_{LR}}$.
The auxiliary layer performed a voxelwise summation of the two channels of $\mathrm{\hat{Y}_{LR}}$
corresponding to left and right lung prediction. The resulting single-channel
image was concatenated
with the background channel of $\mathrm{\hat{Y}_{LR}}$. This produced a two-channel prediction image,
denoted $\mathrm{\hat{Y}_{T}}$, with the channels corresponding to lung vs.
background. During training, supervision was provided at both $\mathrm{\hat{Y}_{LR}}$ and
$\mathrm{\hat{Y}_{T}}$. Equal numbers of human and animal images were
sampled for each batch. Ground truth images were 
denoted $\mathrm{Y_{LR}}$ for labeled images that
distinguished left vs. right lung, and 
$\mathrm{Y_{T}}$
for labeled images that had a single label for total lung.
The loss between
$\mathrm{\hat{Y}_{LR}}$ and $\mathrm{Y_{LR}}$ was computed using only the human
half of the batch, while
the loss between $\mathrm{\hat{Y}_{T}}$ and
$\mathrm{Y_{T}}$ was computed using the entire batch by converting
$\mathrm{Y_{LR}}$ to $\mathrm{Y_{T}}$ for human cases. These two
losses were equally weighted during each training step.

\subsection*{Lobar Analysis}
Lobar segmentations were obtained by using the proposed left and
right lung segmentation as input to the 
FissureNet and LobeNet algorithms, 
which is currently the leading performer in the LOLA11 grand challenge.
No additional training of FissureNet and LobeNet was
performed. Regional lobar analysis was performed using hierarchical
clustering to identify subtypes of COVID-19.

\subsection*{Ablation Study}
To evaluate the contribution of the polymorphic training approach for
lung segmentation, the proposed approach was compared to a
nonpolymorphic model. The nonpolymorphic model only used the human
CT images of COPD for training (i.e., the auxiliary layer and animal
training data were not utilized). Otherwise, there were no differences in
the design or training of the polymorphic and nonpolymorphic models.
A two-way analysis of variance was performed with model type as a
categorical variable and nonaerated lung volume fraction as a
continuous variable, as well as an interaction term.

\section*{Results}

\subsection*{Lung Segmentation}
\begin{figure}[ht]
    \centering 
    \begin{tabular}{@{}c@{\hspace{0.6em}}c@{\hspace{0.6em}}c@{}}
          CT Image & Non Poly & Poly \\
         \includegraphics[trim=5 10 5 10, clip,width=0.33\linewidth]{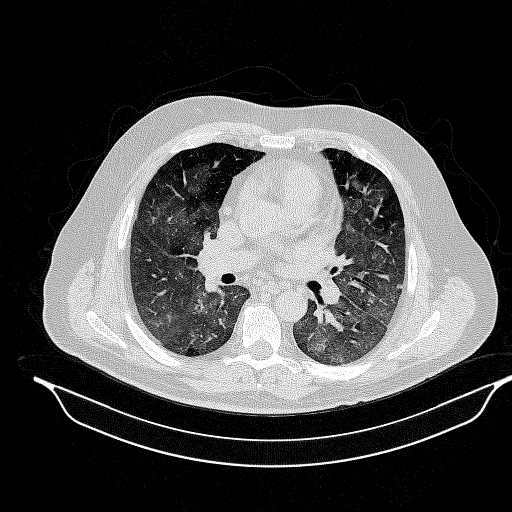} &
         \includegraphics[trim=5 10 5 10, clip,width=0.33\linewidth]{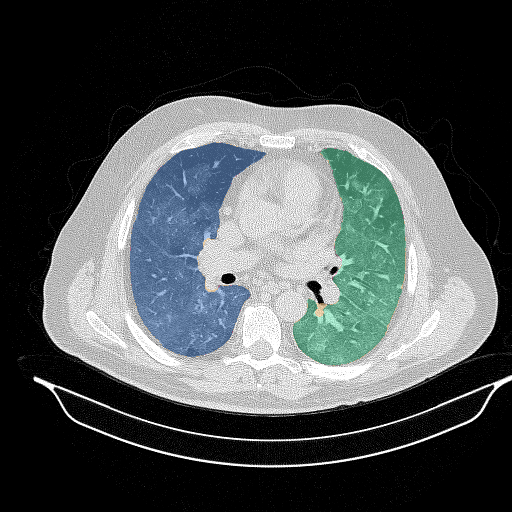} &
         \includegraphics[trim=5 10 5 10, clip,width=0.33\linewidth]{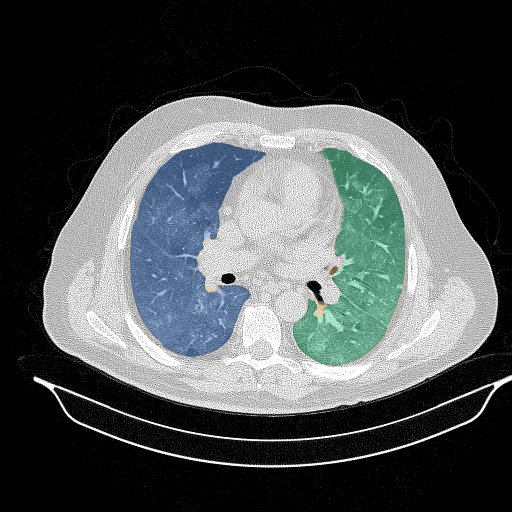} \\
         \includegraphics[trim=5 10 5 10,clip,width=0.33\linewidth]{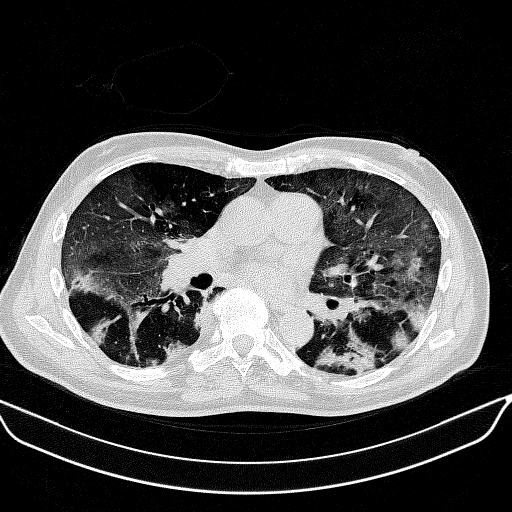} &
         \includegraphics[trim=5 10 5 10,clip,width=0.33\linewidth]{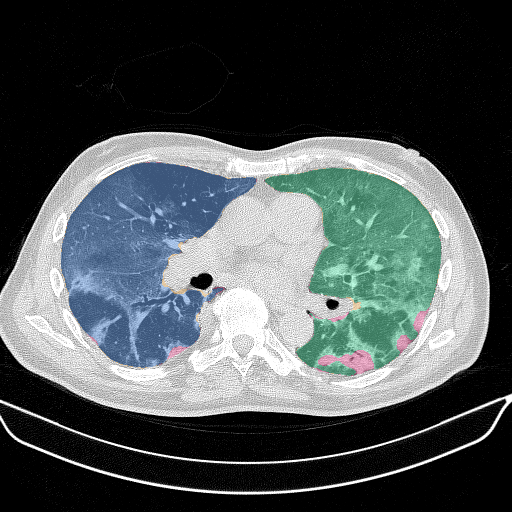} &
         \includegraphics[trim=5 10 5 10, clip,width=0.33\linewidth]{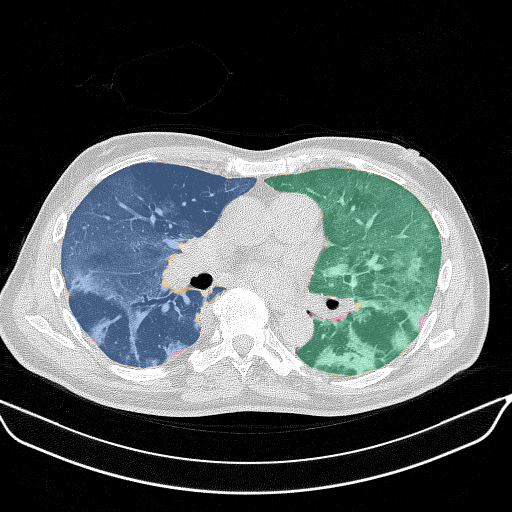} \\
         \includegraphics[trim=5 10 5 10, clip,width=0.33\linewidth]{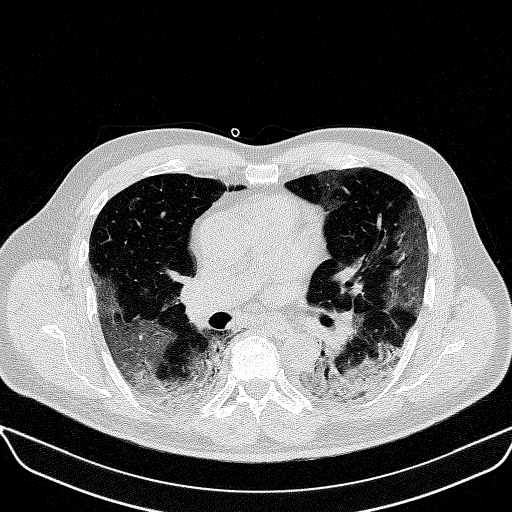} &
         \includegraphics[trim=5 10 5 10, clip,width=0.33\linewidth]{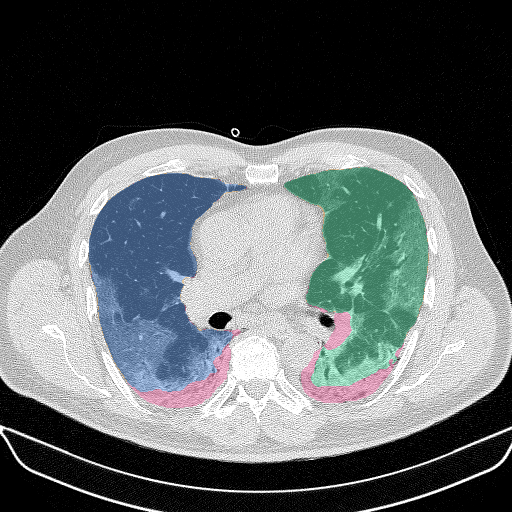} &
         \includegraphics[trim=5 10 5 10, clip,width=0.33\linewidth]{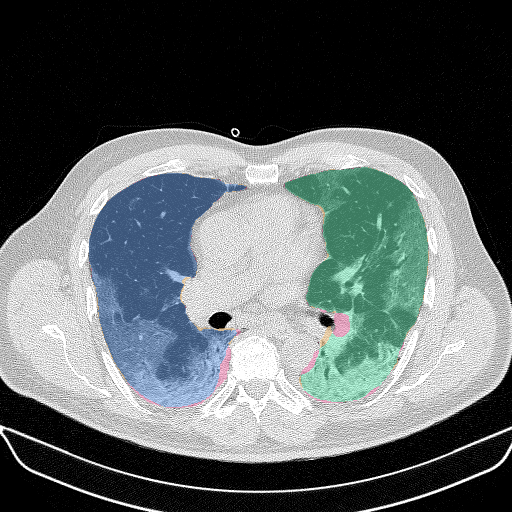} \\
         \includegraphics[trim=5 10 5 10, clip,width=0.33\linewidth]{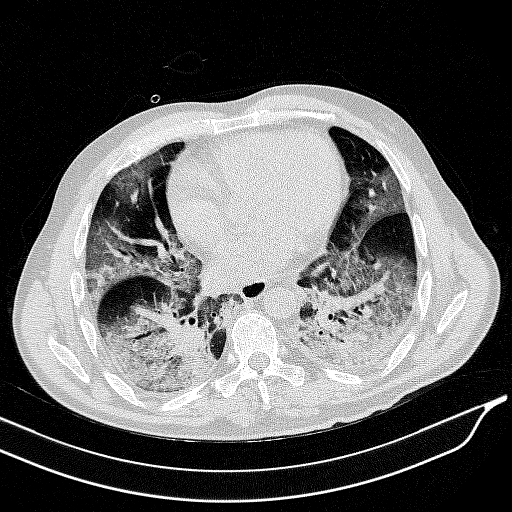} &
         \includegraphics[trim=5 10 5 10, clip,width=0.33\linewidth]{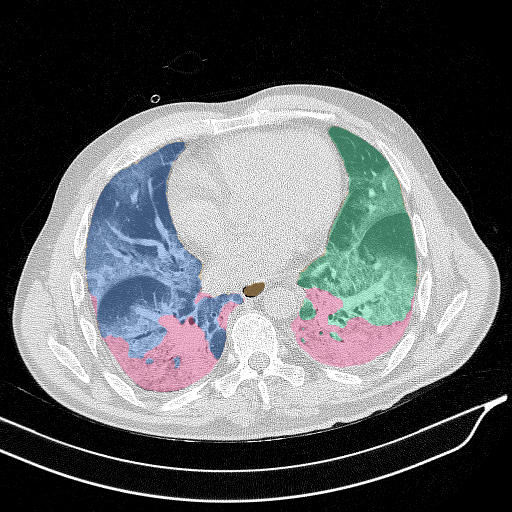} &
         \includegraphics[trim=5 10 5 10, clip,width=0.33\linewidth]{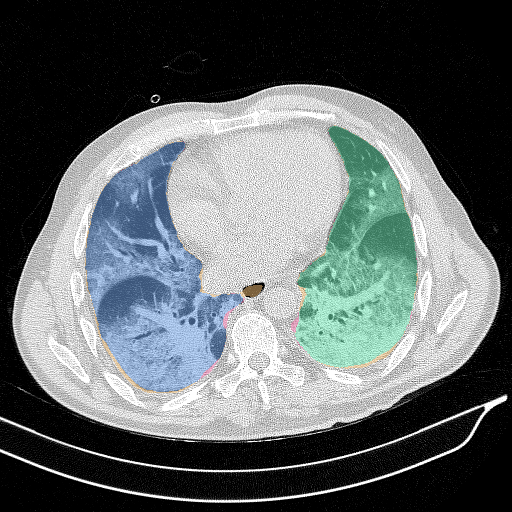} \\
\end{tabular}
     \caption{Axial slices of CT images (left column) and lung segmentation results for the
     nonpolymorphic model (center column) and the polymorphic model (right
     column) algorithms for four COVID-19
     patients (by row). Correctly classified
     voxels are displayed in blue and green for right and left lungs,
     respectively. False negative and false positive voxels are
     illustrated in pink and yellow, respectively.}
     \label{fig:lungslices}
 \end{figure}

\begin{figure}[ht]
     \begin{tabular}{c}
         \includegraphics[width=.9\columnwidth]{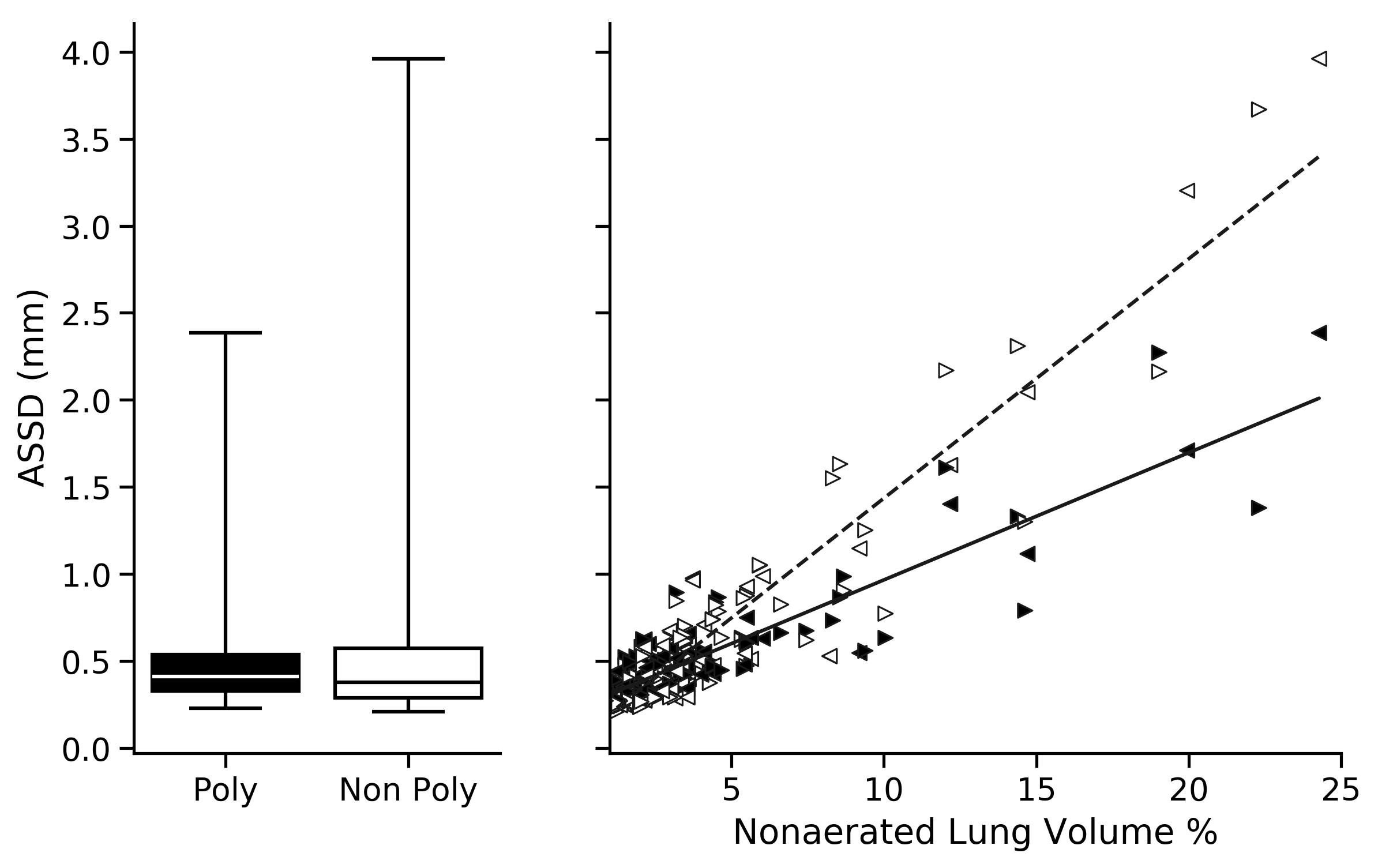} \\
         \includegraphics[width=.9\columnwidth]{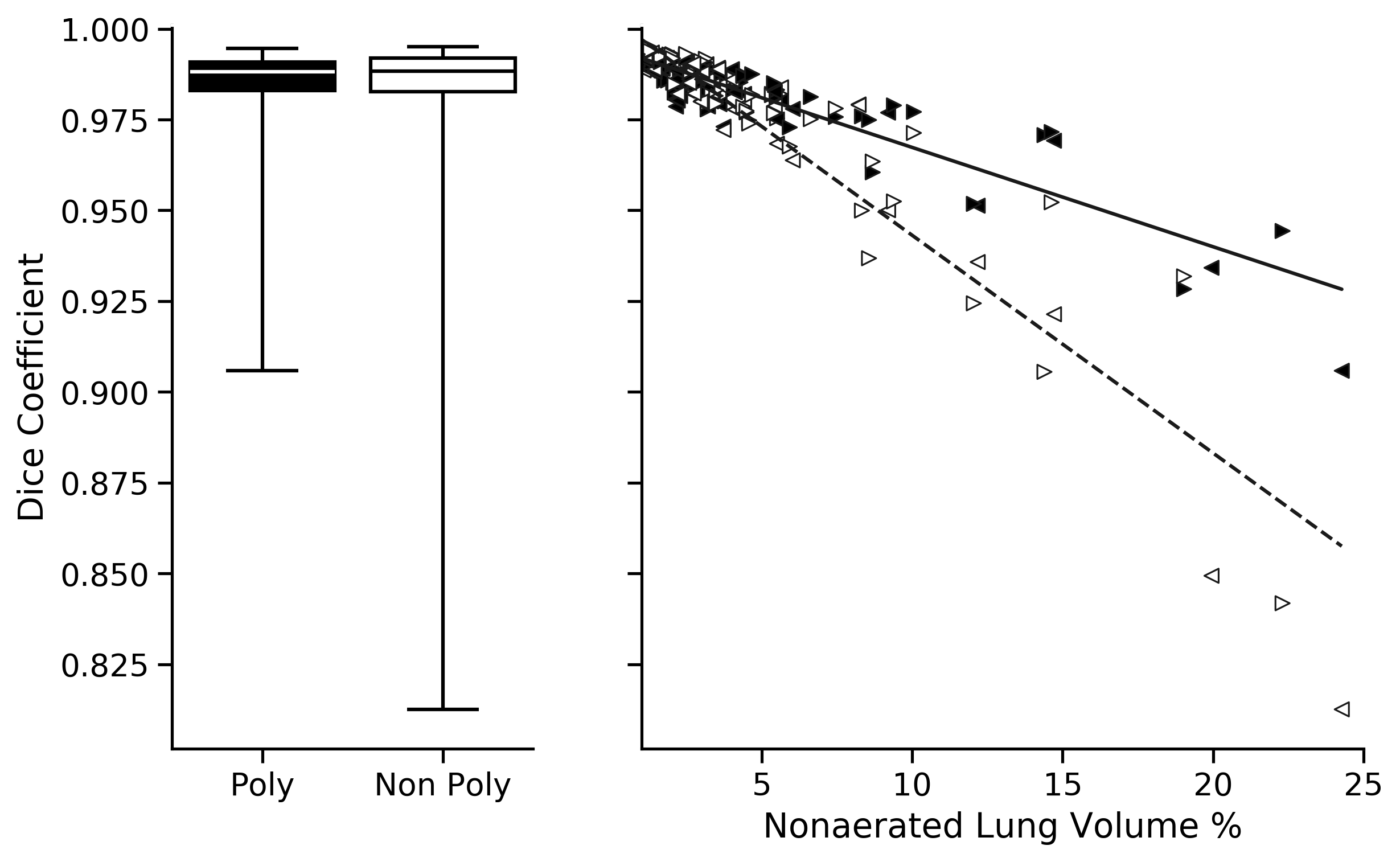}
     \end{tabular}
     \caption{
    Quantitative evaluation of lung segmentation on the COVID
     evaluation dataset ($N=87$). The proposed polymorphic model
     (black)
     is compared to a nonpolymorphic model (white) using ASSD
     and the Dice coefficient. Results are stratified by
     nonaerated lung volume percent in the right panel.
     Left and right lung results are denoted using left- and
     right-facing triangles, respectively (left:
     $\blacktriangleleft\vartriangleleft$, right:
     $\blacktriangleright\vartriangleright$).
      Linear regression for polymorphic (solid) and nonpolymorphic
      (dashed) models revealed significantly different coefficients for
      ASSD in mm $\%^{-1}$ (polymorphic: 0.073, nonpolymorphic:
      0.138, $p<0.001$) and Dice coefficient in $\%^{-1}$ (polymorphic:
      -0.003, nonpolymorphic: -0.006, $p<0.001$).}
     \label{fig:boxplot}
 \end{figure}

Lung segmentation results for the polymorphic and nonpolymorphic
models are shown in Figure~\ref{fig:lungslices}.  Quantitative
evaluation of lung segmentations was performed on CT images by
comparing the segmentations to ground truth manual segmentations.
The Dice coefficient was used to measure volume overlap and the
average symmetric surface distance (ASSD) was used to assess boundary
accuracy. The ASSD and Dice coefficient results for each of the four
evaluation datasets are shown in Table~\ref{tab:results}.  Overall,
on the COVID-19 dataset the polymorphic model achieved an average
ASSD of $0.495\pm0.309$ mm and average Dice coefficient of
$0.985\pm0.011$. By comparison, the nonpolymorphic model achieved an
average ASSD of $0.550\pm0.546$ mm and average Dice coefficient of
$0.982\pm0.024$.  ASSD and Dice coefficient results with respect to
nonaerated lung volume fraction are displayed in
Figure~\ref{fig:boxplot}.  Two-way analysis of variance revealed a
significant interaction between model and nonaerated fraction for
each evaluation metric, indicating that the regression coefficients
with respect to nonaerated fraction were significantly different for
polymorphic vs.  nonpolymorphic models.  
 
 \begin{table}
     \centering
     \caption{Lung segmentation results for polymorphic (Poly) and
     nonpolymorphic (Non-Poly) models. Results are stratified by lung
     (LL: left lung, RL: right lung) and the four evaluation
     datasets. ASSD results are in mm.}
     \begin{tabular}{llcccccccc}
\toprule

     &   & \multicolumn{2}{c}{COPD} & \multicolumn{2}{c}{Cancer} & \multicolumn{2}{c}{IPF} & \multicolumn{2}{c}{COVID-19} \\
     &   & \multicolumn{2}{c}{$N=5986$} &
     \multicolumn{2}{c}{$N=1620$} & \multicolumn{2}{c}{$N=305$} &
     \multicolumn{2}{c}{$N=87$} \\
\midrule
     &   &     LL &     RL &     LL &     RL &     LL &     RL & LL & RL \\
\midrule

    \multirow{2}{*}{ASSD} & Non Poly &  0.339 &  0.300 &  0.355 &  0.485 &  0.478 &  0.500 &    0.514 &  0.586 \\
     & Poly &  0.378 &  0.346 &  0.430 &  0.513 &  0.505 &  0.594 &    0.480 &  0.510 \\
     \midrule
    \multirow{2}{*}{Dice} & Non Poly &  0.990 &  0.992 &  0.990 &  0.987 &  0.985 &  0.985 &    0.982 &  0.982 \\
     & Poly &  0.989 &  0.991 &  0.988 &  0.986 &  0.984 &  0.982 &    0.984 &  0.985 \\
\bottomrule
\end{tabular}

     \label{tab:results}
 \end{table}
 
\subsection*{Lobar Segmentation}
\begin{figure}[ht]
    \centering 
     \begin{tabular}{@{}c@{\hspace{0.6em}}c@{\hspace{0.6em}}c@{}}
         CT Image & PTK & Proposed \\
         
         \includegraphics[width=0.25\linewidth]{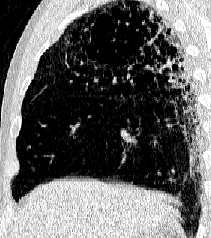} &
         \includegraphics[width=0.25\linewidth]{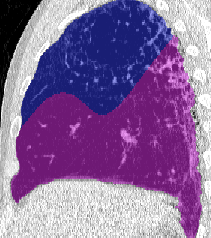} &
         \includegraphics[width=0.25\linewidth]{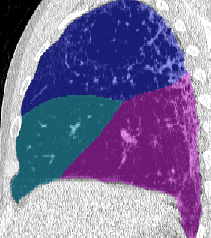} \\
         
         \includegraphics[width=0.25\linewidth]{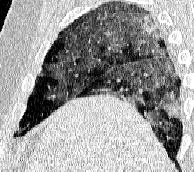} &
         \includegraphics[width=0.25\linewidth]{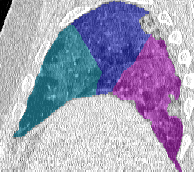} &
         \includegraphics[width=0.25\linewidth]{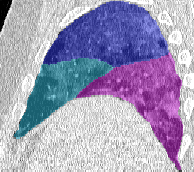} \\
         
         \includegraphics[width=0.25\linewidth]{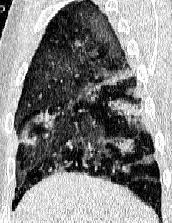} &
         \includegraphics[width=0.25\linewidth]{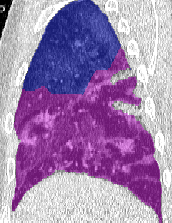} &
         \includegraphics[width=0.25\linewidth]{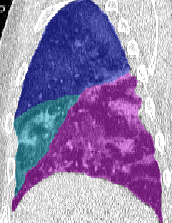} \\
         
         \includegraphics[width=0.25\linewidth]{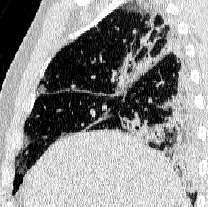} &
         \includegraphics[width=0.25\linewidth]{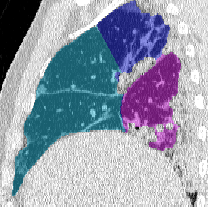} &
         \includegraphics[width=0.25\linewidth]{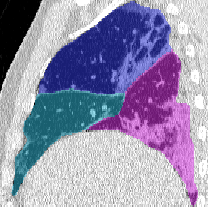} \\
         
\end{tabular}
         \caption{Sagittal slices of CT images (left column) and right lobe segmentation results for the PTK
     (center column) and proposed (right column) algorithms for
     four COVID-19
         patients (by row)}
     \label{fig:lobeslices}
 \end{figure}

\begin{figure}[ht]
    \centering 
     \begin{tabular}{@{}c@{\hspace{0.6em}}c@{\hspace{0.6em}}c@{}}
         CT Image & PTK & Proposed \\
         
         \includegraphics[width=0.25\linewidth]{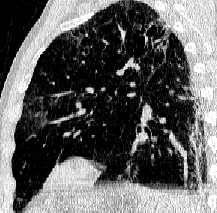} &
         \includegraphics[width=0.25\linewidth]{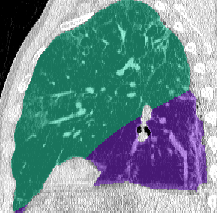} &
         \includegraphics[width=0.25\linewidth]{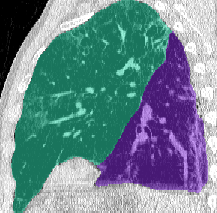}\\
         
         \includegraphics[width=0.25\linewidth]{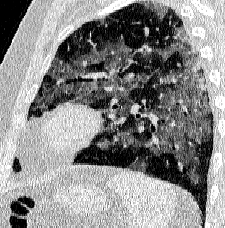} &
         \includegraphics[width=0.25\linewidth]{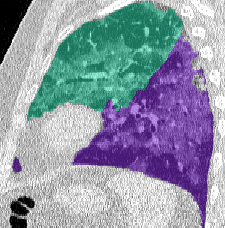} &
         \includegraphics[width=0.25\linewidth]{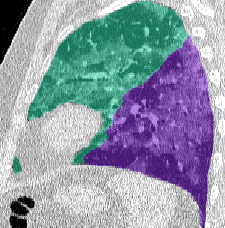}\\
         
         \includegraphics[width=0.25\linewidth]{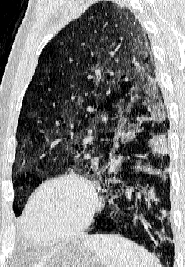} &
         \includegraphics[width=0.25\linewidth]{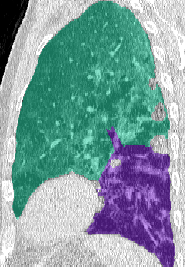} &
         \includegraphics[width=0.25\linewidth]{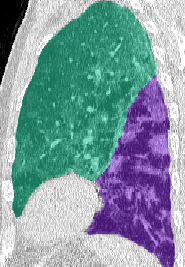}\\
         
         \includegraphics[width=0.25\linewidth]{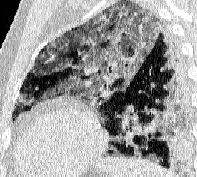} &
         \includegraphics[width=0.25\linewidth]{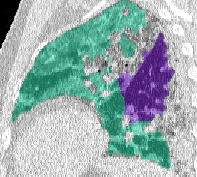} &
         \includegraphics[width=0.25\linewidth]{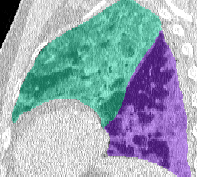}\\
         
\end{tabular}
         \caption{Sagittal slices of CT images (left column) and left lobe segmentation results for the PTK
     (center column) and proposed (right column) algorithms for
     four COVID-19
         patients (by row).}
     \label{fig:lobeslicesleft}
 \end{figure}

\begin{figure}[ht]
\centerline{\includegraphics[width=.9\columnwidth]{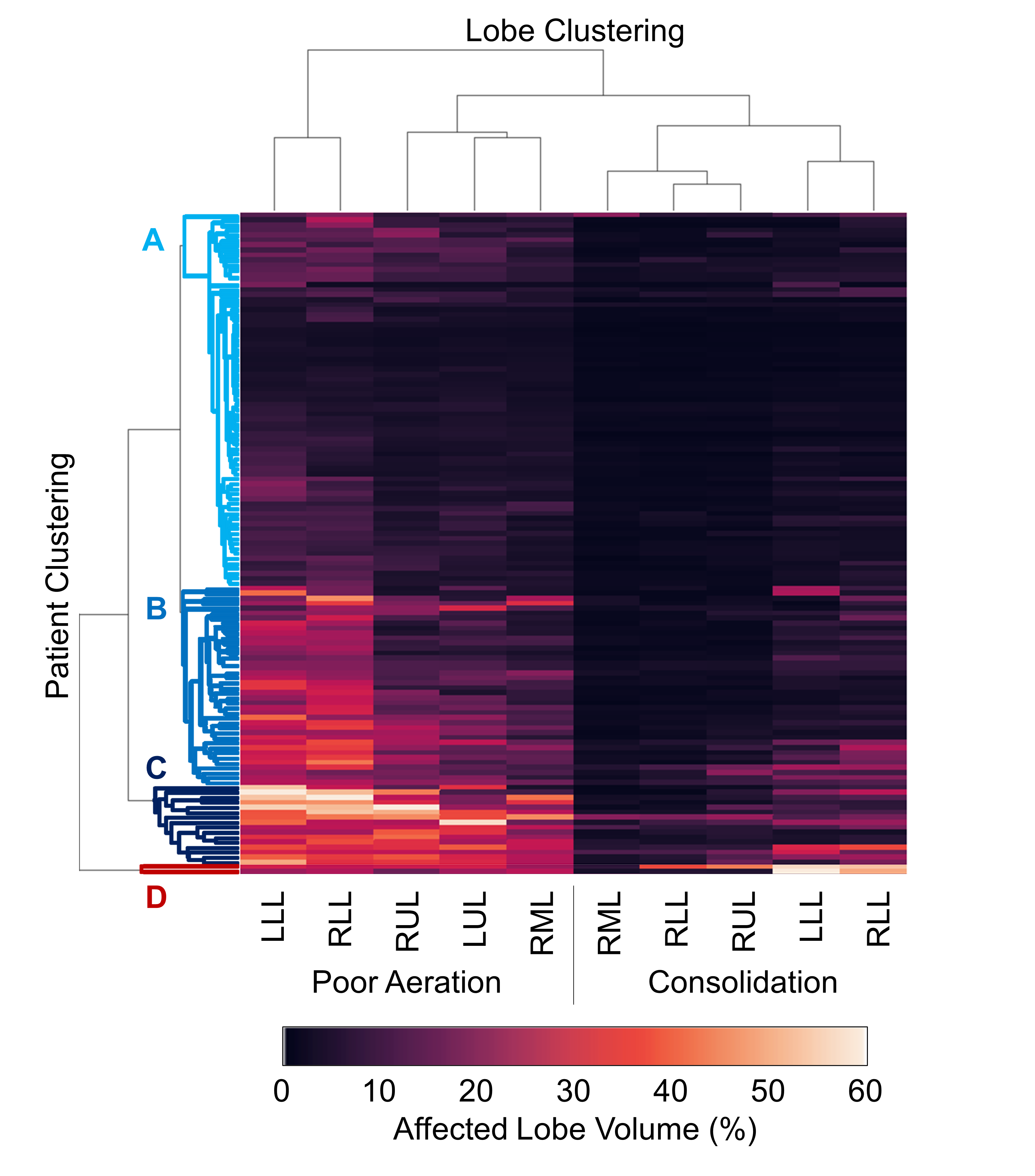}}
    \caption{Hierarchical clustering results showing disease subtypes
    of COVID-19 patients. Each row corresponds to one patient.
    The left five columns show percent of lobe volume with
    poor aeration ($-500 < \mathrm{HU} <
    -100$) and the right five columns show percent
    of lung lobe volume with consolidation ($\mathrm{HU}\geq-100$).
    Poor aeration is used as an approximation of ground glass
    opacities. The dendrogram visualization shows four subtypes of
    patients:
(a) mild loss of aeration
primarily in the two lower lobes without consolidation, 
(b) moderate loss of aeration focused in the two lower lobes with or without
consolidation in lower lobes,
(c) severe loss of aeration throughout all lobes with or without
consolidation, and (d) severe loss of aeration and consolidation throughout all lobes.}
\label{fig:heatmap}
\end{figure}

Lobar segmentation results for the proposed method and PTK are
shown in Figure~\ref{fig:lobeslices} for right lungs and Figure~\ref{fig:lobeslicesleft} for left lungs.
For each image in the COVID-19 dataset (133 images in total), the
lobar segmentation result was used to extract the amount of poor
aeration ($-500 < \mathrm{HU} < -100$) and consolidation
($\mathrm{HU}\geq-100$) in each lobe.  Common phenotypes of COVID-19
affected lungs were identified by hierarchical clustering over the
fraction of poorly aerated and consolidated tissue in each lobe.
Dendrographic analysis in Figure~\ref{fig:heatmap} reveals four
primary clusters of patients that were identified by the hierarchical
clustering: (a) mild loss of aeration primarily in the two lower
lobes without consolidation; (b) moderate loss of aeration focused in
the two lower lobes with or without consolidation in lower lobes; (c)
severe loss of aeration throughout all lobes with or without
consolidation; and (d) severe loss of aeration and consolidation
throughout all lobes. 


\section*{Discussion}
In this study, we proposed and implemented a novel polymorphic training algorithm for lung and lobar segmentation in a fully automated pipeline. The pipeline was independently evaluated on CT scans
of subjects with COVID-19, lung cancer, and IPF - however, no COVID-19,
lung cancer, or IPF scans were utilized for training the CNNs. Additionally, the pipeline was extensively evaluated on CT
scans of patients with COPD. The COVID-19 scans are considered very
challenging cases for lung and lobe segmentation. Peripheral and
diffuse opacities result in little contrast at the lung boundary. In
many cases, the fissure appearance was irregular due to close
proximity of infection.  Furthermore, these are clinical scans with
some cases having slice thickness greater than 3 mm. Fissure
segmentation is especially challenging on such cases. Success of the
proposed algorithm on these cases lends to the generalizability of the
proposed approach.

Out lung segmentation algorithm was quantitatively evaluated
on 7998 CT images, consisting of four distinct pulmonary pathologies. To our knowledge, this is the
most extensive evaluation of a lung segmentation algorithm to date.
The polymorphic and nonpolymorphic models both achieve sub-voxel lung
segmentation accuracy and demonstrate generalizability across
datasets and diseases which were not used for training.  The
polymorphic and nonpolymorphic models achieved similar performance on
COPD, IPF, and lung cancer cases and on COVID-19 cases without
consolidation.  The ablation studied demonstrated that the
polymorphic model was able to accurately segment COVID-19 cases with
severe consolidation, whereas the nonpolymorphic model failed on
such cases.

Gerard et al proposed a transfer learning approach for
lung segmentation in animal images, using a network pre-trained on
human datasets~\cite{gerard2020multi}. This resulted in two networks
that performed well in their respective domains: humans with COPD,
and animals with diffuse opacities.  However, neither network was developed to
performed adequately in the domain of humans with diffuse opacities. In this study, we utilized the human and animal
datasets for training in a combined domain, which led to accurate
performance on human datasets with diffuse opacities and
consolidation (COVID-19). This was achieved using novel polymorphic
training to accommodate both human and animal datasets with different
degrees of label specificity.  The lung module trained only with COPD
datasets (i.e., nonpolymorphic training) performed poorly on COVID-19
cases with consolidation.  By contrast, the fissure and lobar modules
showed high performance despite being trained on COPD datasets
exclusively. 

Our lung segmentation which identifies left and right lungs can be
used as input to the LobeNet algorithm to achieve lobar segmentation.
The lobar segmentations can be used to quantify involvement of disease at
the lobar level, and thus may identify clusters of patients with similar
phenotypes indicative of disease stage or prognosis. Pan et al. reported predominant lower lobe involvement in early disease
that progresses to all lobes at the peak of disease
severity~\cite{pan2020time}. Inui et al. reported similar findings
in the Diamond Princess cohort and also found that 83\% of
asymptomatic patients have more ground glass opacities than consolidation compared to
only 59\% of symptomatic patients~\cite{inui2020chest}. The four
\textit{quantitatively} identified clusters in our study match the
results of \textit{qualitative} scoring performed by radiologists in
these studies~\cite{pan2020time,inui2020chest}.  Cluster (a) is
similar to early disease phenotype with predominantly ground glass
opacities in the
lower lobes; cluster (d) is similar to peak disease phenotype with
large amounts of consolidation and ground glass opacities in all lobes; and clusters
(b) and (c) may represent transitional phenotypes. Clinical
information could be used to validate this analysis. Huang at el.
performed a similar lobar analysis using a deep learning approach and
also reported increasing
opacification with disease progress. However, they did not show
lobar segmentation results in a manner that allows us to qualitatively
assess their accuracy~\cite{huang2020serial}.

Our computational pipeline required an average of 2.5 minutes to run on a GPU. By
comparison, manual segmentation of lungs and lobes takes several
hours, which is not feasible in clinical settings.  Our
approach thus allows regional quantification of disease at
the lobar level, which would otherwise not be possible in such a
short time frame.  Lobar characterization of disease involvement may
also assist in identifying subtypes of COVID-19 for treatment
stratification.

A limitation of the current work is lack of comparison to other lung
segmentation methods. Given this is the first attempt to
handle training data with different levels of specificity, other
comparisons would be limited to training on only the COPD dataset.
This would not be an appropriate comparison for evaluation on COVID-19 cases,
as demonstrated by the ablation study in this work.  Another
limitation is the number of COVID-19 cases available, making it
difficult to draw conclusions from the regional analysis. We
only proposed a type of analysis that can be performed, and did not
make any conclusions regarding disease prognosis and stratification. In this work,
polymorphic training approach was applied to identifying left vs.
right lung. However, this approach could be generalized to other
problems with hierarchical labels. A natural extension of this work
is to apply the polymorphic training to lobes, which can be explored in the future.

\section*{Conclusion}

In summary, we have demonstrated a robust deep learning pipeline for lung and
lobar segmentation of CT images in patients with COVID-19, without requiring
previously segmented COVID-19 datasets for training. A novel polymorphic
algorithm was proposed to accommodate training data with different
levels of label specificity. Our approach
accurately segmented lungs and lobes across various pulmonary
diseases, including challenging cases with diffuse consolidation seen in
critically-ill COVID-19 patients.  Automated and reliable segmentation is
critical for efficient and objective quantification of infection from CT
images, and may be valuable for identifying subtypes and monitoring progression
of COVID-19.

\clearpage 



\section*{Acknowledgements}


We thank Parth Shah, Shiraz Humayun, Paolo Delvecchio, Debanjan Haldar, Gayatri Maria Schur, Noah Mcqueen who have worked on the manual segmentation.
We thank Dr. Chang Hyun Lee of Seoul National University and Dr. Kyung Min Shin of Kyungpook National University School of Medicine, Daegu, South Korea for contributing CT scans of COVID-19 patients.
We thank Guido Musch, Ana Fernandez-Bustamante, and Brett A. Simon for providing ovine
animal datasets.

This work was supported in part by NIH grants R01-HL142625 and R01-HL137389, and by a grant from the Carver Charitable Trust.
This work was supported by the Office of the Assistant
Secretary of Defense for Health Affairs through the Peer-Reviewed
Medical Research Program under Award No. W81XWH-16-1-0434.  Opinions,
interpretations, conclusions, and recommendations are those of the
authors and are not necessarily endorsed by the Department of
Defense.

We thank the COPDGene investigators for providing the
human image datasets used in this study. The COPDGene study is
supported by NIH grants R01 HL089897 and R01 HL089856.


\section*{Author contributions statement}



S.E.G and J.M.R. made substantial contributions to the conceptualization of the work.
S.E.G., Y.X, K.T.M., E.R., D.I., G.B., M.C., J.G. were involved with acquisition, analysis, and/or interpretation of data.
S.E.G. wrote the new software used in this work.
S.E.G. drafted the manuscript. J.H., J.M.R., D.W.K., E.A.H. substantively revised the manuscript.
All authors reviewed and approved the submitted manuscript.

%
%
%
%
%

\end{document}